\documentclass[aps,pra,reprint]{revtex4-2}
\usepackage{blindtext}
\usepackage{amsmath}
\usepackage{amssymb}

\usepackage{setspace}
\usepackage{xcolor}
\usepackage{transparent}
\usepackage{comment}
\usepackage{graphicx}

\usepackage{todonotes}

\usepackage{pgf}
\usepackage{layouts}

\usepackage[font=small,justification=Justified]{caption}

\newcommand{\ah}{\hat{a}}
\newcommand{\bh}{\hat{b}}
\newcommand{\ddh}{\hat{d}}

\newcommand{\affilUKON}{Department of Physics, University of Konstanz, 78457 Konstanz, Germany.}

\graphicspath{{images/}}

\begin{document}
\title{Quantized nonlinear kink movement through topological boundary state instabilities}
\author{Markus Bestler}
\author{Oded Zilberberg}
\affiliation{\affilUKON}

\begin{abstract}
    Thouless pumping is a paradigmatic example of topologically protected, directed transport in linear systems. Recent extensions to nonlinear pumps often overlook the need to reassess the conventional framework of linear topology. In this work, we study a nonlinear dimer-chain model that exhibits quantized transport of kinks under a periodic modulation of a pumping parameter. Crucially, linear excitations in the system map to a Rice-Mele model and display topological boundary modes localized at these kinks. Using methods from nonlinear dynamics, we show that instabilities in these boundary modes are the driving mechanism behind the observed kink motion. While the transport resembles that of a linear Thouless pump, it cannot be fully captured by conventional topological indices. Instead, the behavior is more akin to a topological ratchet: robust, directional, and reproducible, yet fundamentally nonlinear. Furthermore, by introducing multiple pumping parameters, we demonstrate fine control over multiple kink trajectories, as well as soliton motion, suggesting applications in information transport. Our results unify concepts from linear topology and nonlinear dynamics to establish a framework for quantized transport in nonlinear media.
\end{abstract}
\maketitle

The quantum Hall (QH) effect is a fundamental phenomenon in two-dimensional (2D) materials subject to a perpendicular magnetic field~\cite{cage2012quantum,ozawa2019topological}, marked by quantized transverse transport in response to an in-plane force. This quantization is governed by a topological invariant, the Chern number, which characterizes the system’s bulk spectrum~\cite{laughlin1981HallConductivity,TKNN_Thouless1982,bernevig2013topological,frohlich2023gauge}. A dynamical analogue thereof is the Thouless pump~\cite{Quantization_Thouless1983,kraus2012topological,lohse2016thouless,citro2023thouless}, where a 2D QH system is mapped onto a one-dimensional (1D) system with a time-dependent parameter replacing one quasimomentum. Here, the Chern number counts both the number of particles transported and the number of localized boundary states crossing the gap per pump cycle. This correspondence between bulk topology and boundary phenomena is referred to as the pump’s bulk-boundary correspondence~\cite{kraus2012topological,BulkEdgePumping_Hatsugai2016}. Originally proposed for electronic systems~\cite{Quantization_Thouless1983}, Thouless pumps have since been realized in various platforms, including ultracold atoms~\cite{lohse2016thouless,citro2023thouless} and photonic platforms~\cite{kraus2012topological,FibonacciPump_Zilberberg2015}, with extensions to higher-dimensional settings~\cite{kraus2013four,lohse2018exploring,zilberberg2018photonic,petrides2018six,petrides2020higher}.

Interactions in QH systems can produce exotic correlated states, most notably the fractional QH effect, characterized by a fractionally quantized transverse current in response to an in-plane force~\cite{stormer1999fractional,halperin2020fractional}. A dynamical analogue exists in this regime too~\cite{taoThouless1983}: placing the 2D fractional QH system on a cylinder and shrinking its circumference maps it to a 1D limit where, e.g., the fractional Laughlin ground state becomes a charge density wave. This is the Tao-Thouless limit, with the pumping parameter manifesting as the magnetic flux threaded through the cylinder. Contemporary realizations and proposals for interacting and fractional pumping include ultracold atoms~\cite{Esslinger2023BreakdownInHubbardThoulessPump,Viebahn2024InteractionThoulessPump, jürgensen2025multibandfractionalthoulesspumps} and superconducting junction arrays~\cite{weisbrich2023fractional}.

The fractional QH effect and the Tao–Thouless limit arise in strongly interacting many-body systems. In contrast, recent work has focused on weakly interacting bosonic systems, where mean-field nonlinear topology has become a central focus~\cite{Schneider2025UFNHSE,Bestler2025SkinModesViaParametric}. In this setting, bosonic nonlinear Thouless pumps have emerged as key examples~\cite{Aidelsburger2018InteractingBosonicRiceMele,jurgensen2021quantizedPumping,mostaan2022quantized,jurgensen2023quantized,Ravets2025KerrResonatorThoulessPump}. Unlike linear systems, nonlinear systems can support exponentially many stable steady states. A key question is whether topological features of the linear model extend to the nonlinear regime, where phase transitions between steady states may occur via dynamical instabilities during the pump cycle~\cite{Tuloup2023InstabilitiesInPumpCycle}. Experimental and theoretical work suggests that the displacement of solitons—nonlinear analogues of eigenstates—can be governed by the Chern number of the linear band from which they bifurcate, mirroring the behavior seen in the linear case~\cite{jurgensen2021quantizedPumping,fangwei2022ThoulessPumping,mostaan2022quantized,jurgensen2023quantized,Rechtsman2022SolitonTheory}. However, recent studies question this, proposing that nonlinearity may play a more intrinsic role in determining the system's pumping behavior~\cite{Tao2024nonlinearityinducedthoulesspumpingsolitons,Tao2025NonlinearityInducedFractionalTopoPump,Fleischhauer2025FractionalTopoPump}. All of these works describe nonlinear Thouless pumps via bulk properties and assume that adiabaticity still holds throughout the pump cycle.

In this work, we demonstrate that the principle of bulk-boundary correspondence offers an alternative perspective to nonlinear Thouless pumps. Specifically, we investigate how instabilities of topological boundary states give rise to a pumping-like mechanism in a nonlinear Rice–Mele model. To this end, we consider a dimerized tight-binding chain of bistable sites with attractive couplings and onsite potentials, both modulated by a pumping parameter. Analyzing linearized fluctuations around a ground state reveals behavior similar to the linear Rice–Mele model, including a bulk-boundary correspondence governed by the Chern number. However, when initialized in a first excited state, the system forms two domains separated by a boundary that discontinuously displaces as the pumping parameter advances. We demonstrate that this motion results from a pressure-induced instability of a topological boundary mode at the domain interface and provide an analytical model capturing this effect. Importantly, we show that domain-wall transport is not directly determined by the Chern number of the linearized system, but rather depends on the chosen pumping trajectory. Finally, we extend our analysis to models with multiple pumping parameters and highlight the high tunability of domain wall dynamics in such systems.

\begin{figure}[tb]
	\centering
	\includegraphics[width=1\columnwidth]{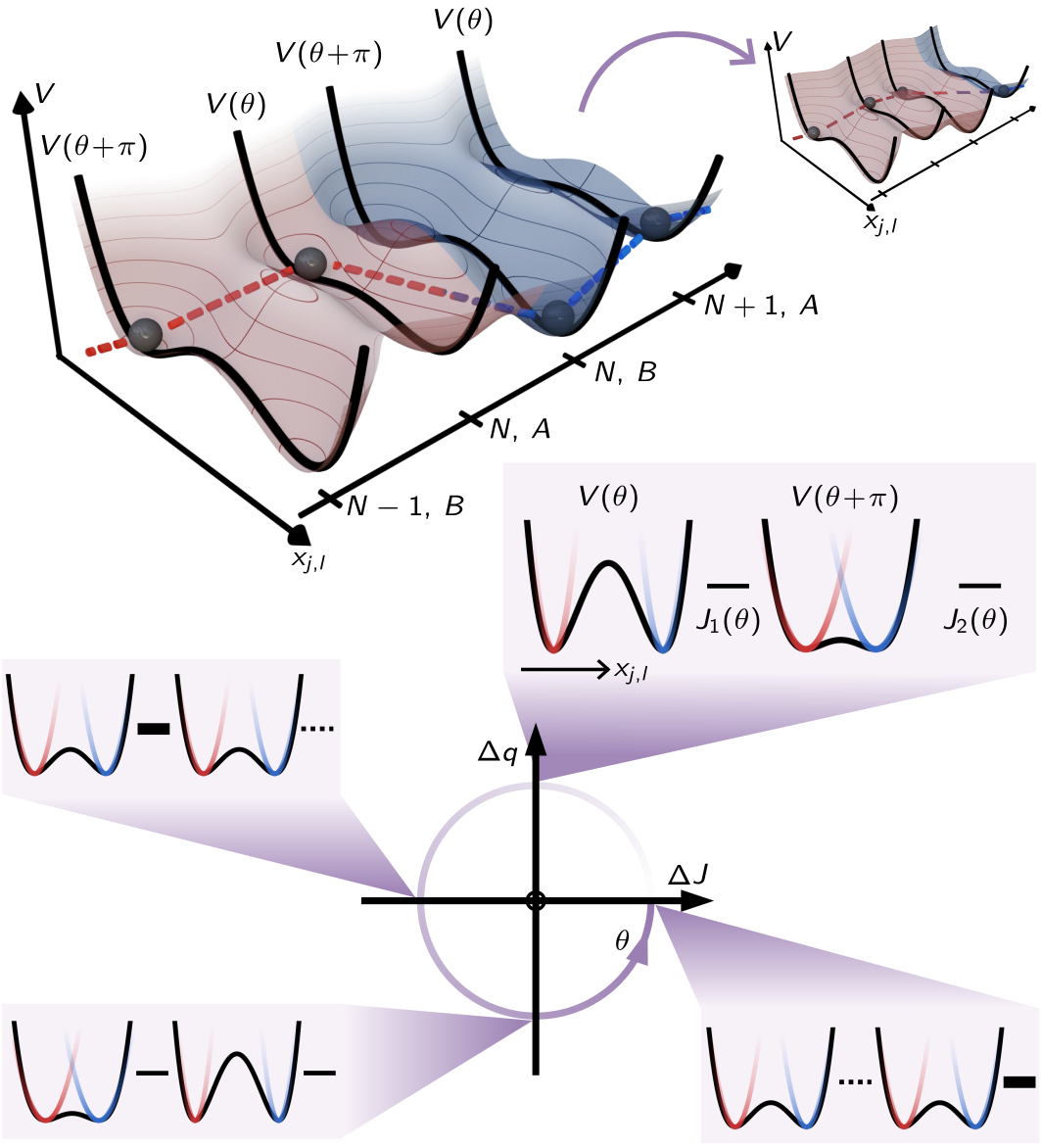}
	\caption{Sketch of the model [cf.~Eq.~\eqref{eq:HamiltonianFullSystem}]. 
     \textbf{a}, A quasi-1d egg-carton potential forming a chain of double-wells with an effective two-site unit cell (sites $A$ and $B$) that repeats along the position $N$ of the chain. In each double-well, a single particle resides subject to an effective 1D potential $V(\theta)$ (black lines). The particles at different sites attract each other with $\theta$ dependent intra- or inter-cell coupling strengths (red and blue dashed lines). Here, the system is configured in a first-excited state with two domains (red and blue) separated by a kink. Advancing the pumping parameter leads to the propagation of this kink. \textbf{b}, Illustration of the double-well potentials in the unit cell and the couplings throughout the pump cycle parametrized by $\theta$ in the plane of potential $\Delta q$ and coupling $\Delta J$ imbalances. Bold, normal and dashed lines mark high, medium and low coupling strengths.}
	\label{Fig:1} 
\end{figure}

We consider a chain with a two-site unit cell with bimodal sites (see Fig.~\ref{Fig:1}\textbf{a})
\begin{align}\label{eq:HamiltonianFullSystem}
    &H(\theta)=\sum_j\bigg[\sum_{l=A,B} \Big(\frac{1}{2}\hat p_{j,l}^2-\frac{q_l(\theta)}{2} \hat x_{j,l}^2+\frac{\lambda}{4}\hat x_{j,l}^4\Big)
    \\
    &+\frac{J_1(\theta)}{2}\left(\hat x_{j,A}-\hat x_{j,B}\right)^2
    +\frac{J_2(\theta)}{2}\left(\hat x_{j,B}-\hat x_{j+1,A}\right)^2\bigg]\, ,\nonumber
\end{align}
where $\hat x_{j,l}$ ($\hat p_{j,l}$) is the position (momentum) operator at site $l\in\{A,B\}$ in the $j$-th unit cell. The bimodality at each site is described by double-well potentials with a constant quartic term $\lambda$ and pumping parameter $\theta$-dependent quadratic terms with
$q_A(\theta) = q_B(\theta+\pi)>0$. The attractive coupling terms are also $\theta$-dependent, with the relation $J_1(\theta) = J_2(\theta+\pi)>0$ between the intra- and inter-cell coupling strengths. In the following, we parametrize our model using
$q_A(\theta) = 1 + \frac{\Delta q(\theta)}{2}$,
$q_B(\theta) = 1 - \frac{\Delta q(\theta)}{2}$,
$J_1(\theta) = \kappa\left(1 - \frac{\Delta J(\theta)}{2}\right)$, and
$J_2(\theta) = \kappa\left(1 + \frac{\Delta J(\theta)}{2}\right)$, with potential and coupling imbalances $\Delta q$ and $\Delta J$, respectively, and an overall coupling constant $\kappa$.
As such, the pumping parameter $\theta$ defines a trajectory in the $\Delta J$–$\Delta q$ plane. For circular trajectories around the origin, we use $\Delta q= \sin(\theta)$ and $\Delta J=\cos(\theta)$ corresponding to counterclockwise winding, see Fig.~\ref{Fig:1}\textbf{b}.

\begin{figure*}[!htb]
	\centering
	\includegraphics[width=1\textwidth]{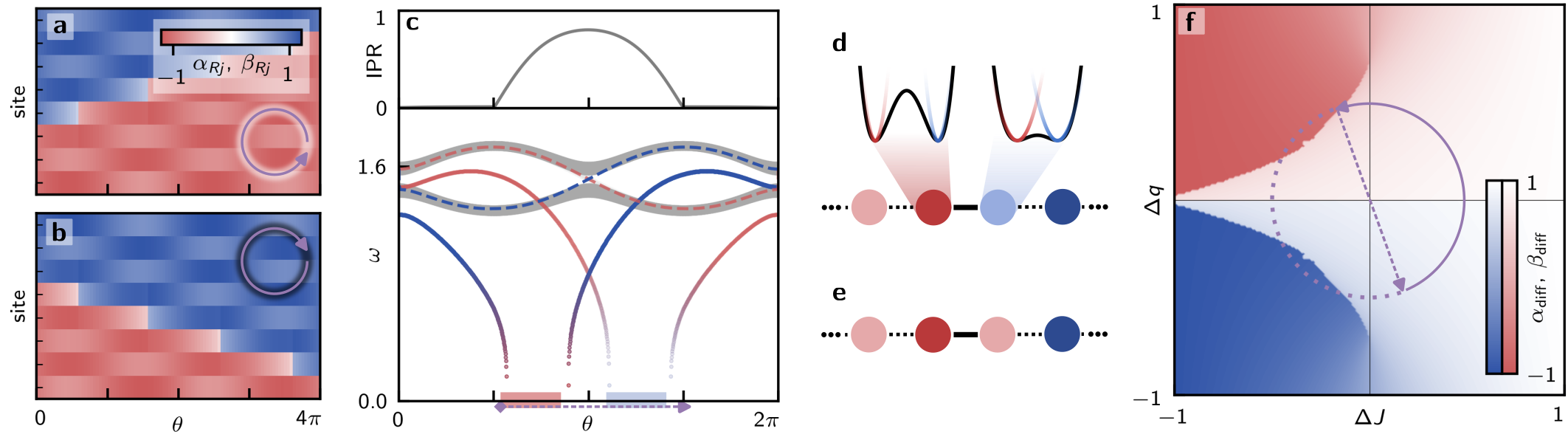}
	\caption{Quantized kink motion via topological boundary mode instabilities for $\lambda=1$ and $\kappa=0.15$ [cf.~Eq.~\eqref{eq:HamiltonianFullSystem}]. 
    \textbf{a,} Displacement within each double well $\alpha_{Rj}\equiv\mathrm{Re}(\alpha_j)$ and $\beta_{Rj}\equiv\mathrm{Re}(\beta_j)$ for an 8-site chain initialized with a central kink, along two counterclockwise circular pump cycles $\theta \in [0, 4\pi]$ [cf.~Fig.~\ref{Fig:1}\textbf{b}]. 
    \textbf{b,} Same as \textbf{a}, but with a clockwise pump direction, $\theta \rightarrow -\theta$. 
    \textbf{c,} Bottom: Quasiparticle bulk bands (gray) on top of a single-domain background for a 100-site chain, as a function of $\theta$ [cf.~Eq.~\eqref{Eq:Rice-MelePart}]. Boundary modes cross the gap for open boundary conditions on the right and left interface to the vacuum (red and blue dashed lines, respectively). The modified boundary mode energies from Eq.~\eqref{eq:ModifiedFrequency} are shown as a solid red (blue) line turning blue (red) at lower energy. Regions where the boundary mode becomes unstable are marked on the x-axis in blue and red, respectively. A change in the system's termination upon reaching an instability corresponds to $\theta \rightarrow \theta+\pi$ (dotted purple arrow). Top: Inverse participation ratio (IPR) for the boundary mode near the nonlinear kink, where IPR$\rightarrow 1$ (IPR$\rightarrow 0$) for localized (delocalized) modes.
    \textbf{d,} and \textbf{e,} Respective schematics of the kink region before and after a boundary mode instability. Pale and solid dots represent the two sublattices, the color indicates the corresponding domain.
    \textbf{f,} Instability phase diagram in the $\Delta J$-$\Delta q$ plane for an 8-site system with initial kink in the center and $\alpha_{\mathrm{diff}}=\mathrm{Re}(\alpha_N/\alpha_0)$, analogous for $\beta_{\mathrm{diff}}$. Red (blue) regions indicate instabilities of the blue (red) domain. The purple circle denotes the pump cycle in \textbf{a} and \textbf{c}, where the transition $\theta\rightarrow\theta+\pi$ upon reaching an instability makes the trajectory jump (dashed arrow); the dotted parts of the pump cycle are never reached.}
	\label{Fig:2} 
\end{figure*}

The ground state of our system has a $\mathbb{Z}_2$ symmetry corresponding to all sites residing either in the left or in the right well, i.e., either $x_{j,l}\equiv\langle \hat{x}_{j,l} \rangle < 0$ or $x_{j,l} > 0$ for all $j$ and $l$. Thus, the system minimizes the coupling energy and forms a single domain. In this work, instead, we focus on the dynamics of excited states; we begin with first-excited states consisting of two distinct domains separated by a single domain wall, also called kink in the following. In each domain, all sites reside uniformly within one of the two double-well configurations, see Fig.~\ref{Fig:1}\textbf{a}. Our main result is: When slowly changing the model’s parameters via a periodic variation of $\theta$ (cf.~Fig.~\ref{Fig:1}\textbf{b}), the kink propagates in a quantized manner. To show this numerically, we first express the position and momentum of each site with a semiclassical mean-field ansatz as    
$\hat x_{j,A} = 1/\sqrt{2} \left( (\alpha_{j} + \alpha_{j}^*) + \epsilon (\ah_{j}^{\phantom \dag} + \ah_{j}^\dagger) \right)$  and  
$\hat p_{j,A} = i/\sqrt{2} \left( (\alpha_{j}^* - \alpha_{j}) + \epsilon (\ah_{j}^\dagger - \ah_{j}^{\phantom \dag}) \right) $,  
where  $\alpha_{j}$ is the coherent state on site $A$ in cell $j$, and $\ah_{j}$ annihilates a quasiparticle excitation on top of that semiclassical state, with a small fluctuation parameter $\epsilon$. For $\hat x_{j,B}$ and $\hat p_{j,B}$ on the $B$ sites, the same expressions hold with $\alpha_{j},\ah_j\rightarrow\beta_j, \bh_j$. Starting from a two-domain state (see Fig.~\ref{Fig:1}\textbf{a}), we find a local mean-field energy minimum [Eq.~\eqref{eq:HamiltonianFullSystem} with $\epsilon=0$] for discretized steps of $\theta$; At each step, we use the resulting ground state to seed the next step. This leads to nonlinear kink movement through the chain, see Figs.~\ref{Fig:2}\textbf{a} and \textbf{b}. In the following, we show that this quantized nonlinear pumping arises from instabilities in the excitations on top of the two domains, particularly of their topological boundary modes. These instabilities, unique to nonlinear systems, drive phase transitions between first-excited states, enabling the kink propagation. The resulting scheme is inherently nonlinear and resembles a ratchet.


In order to eventually predict instabilities of topological boundary modes, we now analyze the excitations ($\epsilon\neq 0$) on top of the semiclassical steady state. For clarity, we first focus on the ground state (one domain, no kinks). Inserting the mean-field ansatz into Eq.~\eqref{eq:HamiltonianFullSystem}, omitting constant energy contributions, and retaining terms up to quadratic order in $\epsilon$, we obtain the linear excitation Hamiltonian
\begin{align}
    \label{Eq:Rice-MelePart}
    H_{\rm fl} = \sum_j & \bigg[\delta_A(\theta) \ah_j^\dag \ah_j^{\phantom \dag} + \delta_B(\theta) \bh_j^\dag \bh_j^{\phantom \dag}
    \\
    &-\frac{J_1(\theta)}{4}\ah_j^\dag \bh_j^{\phantom \dag} - \frac{J_2(\theta)}{4}\bh_j^\dag \ah_{j+1}^{\phantom \dag} +h.c.\bigg]+H_{\rm sq}\, ,\nonumber
\end{align}
which has the form of an effective Rice-Mele model~\cite{RiceMeleModel_Rice1982} with modulated dimerized hopping and on-site potential  $\delta_A = \frac{1}{4} \left( 1 - q_A(\theta) + J_1(\theta) + J_2(\theta) + \frac{3}{2}\lambda (\alpha + \alpha^*)^2 \right)$ depending on the mean-field background. The potential $\delta_B$ takes a similar form with $A,\alpha\rightarrow B,\beta$. The squeezing term $H_{\rm sq}$ acts as a perturbation and shares a similar $\theta$-dependent structure, see Supporting Information. Hence, excitations on top of the ground state experience a variant of the Rice-Mele model for topological pumps~\cite{Quantization_Thouless1983,lohse2016thouless}. Using a Bogoliubov transform~\cite{xiao2009theorytransformationdiagonalizationquadratic}, we find the $\theta$-dependent eigenmodes of $H_{\rm fl} = \sum_i \omega_i \ddh_i^\dagger \ddh_i^{\phantom \dag}$, with eigenenergies $\omega_i$ and quasiparticles annihilation (creation) operator $\ddh_i^{\phantom \dag}$ ($\ddh_i^\dag$), see  Fig.~\ref{Fig:2}\textbf{c}. Similar to the Rice-Mele model, the bands of $H_{\rm fl}$ exhibit nontrivial bulk topology characterized by the pump's Chern number~\cite{ozawa2019topological}, see Supporting Information. The lower (upper) band in Fig.~\ref{Fig:2}\textbf{c} carries Chern number $-1$ ($+1$). Upon opening the boundary of the one-domain system, localized topological modes cross the gap during the pump cycle, due to the pump's bulk-boundary correspondence~\cite{kraus2012topological,Hatsugai2016BulkBoundary}, see Fig.~\ref{Fig:2}\textbf{c}.

We now analyze quasiparticles on top of a two-domain excited state, cf.~Fig.~\ref{Fig:1}\textbf{a}. We perform the same procedure as for Eq.~\eqref{Eq:Rice-MelePart}, but use mean-field solutions corresponding to a two-domain state, where all sites in the left (right) domain occupy the left (right) well, with a kink between the $A$ and $B$ sites of the $N$-th unit cell. Far from the kink, the mean-field background matches that of the respective ground states [cf.~Eq.~\eqref{Eq:Rice-MelePart}]. As such, each domain acts as a topological pump for excitations. Crucially, the kink serves as a boundary between two distinct pumps. Unlike a boundary with the vacuum (cf.~Fig.~\ref{Fig:2}\textbf{c}), the kink behaves as an active boundary, as the domains exert additional force on each other. This implies a pressure build-up at the inter-domain boundary, which modifies the eigenenergies of excitations residing there, see Supporting Information. We show in the following, that a topological excitation mode localizing at this inter-domain boundary becomes unstable, i.e., its linearized eigenenergy turns imaginary. Once the system reaches this instability, it rearranges into a new stable two-domain state where the kink is shifted by one site (see Figs.~\ref{Fig:2}\textbf{a}, \textbf{b}, \textbf{d}, \textbf{e}) and no topological mode remains frustrated by the pressure.

Our hypothesis is that the direction of the kink’s motion is determined by which domain \textit{first} hosts a topological boundary mode that localizes at the kink and becomes unstable. Consequently, it suffices to consider an effective model for a single domain under a boundary pressure force (see Supporting Information):
\begin{align}
    \tilde{H}_{\rm fl}=&
    \epsilon^2 H_{\rm fl}
    +\frac{\epsilon^3\lambda}{4}(\alpha_N+\alpha_N^*) (\ah_N^{\phantom \dagger}+\ah_N^\dagger)^3 + \frac{\epsilon^4\lambda}{16}(\ah_N^{\phantom \dagger}+\ah_N^\dagger)^4\nonumber\\
    & - \epsilon J_1(\theta)(\beta_{N}^{\phantom *}+\beta_{N}^*)(\ah_N^{\phantom \dagger}+\ah_N^\dagger)\, .\label{eq:PreBogoliubovFluctuationHamiltonian}
\end{align}
We assume the kink lies between the $A$ and $B$ sites of the $N$-th unit cell, see Fig.~\ref{Fig:1}\textbf{a}. Starting from the single-domain model~\eqref{Eq:Rice-MelePart}, we include only the coupling to the mean-field amplitude $\beta_N$ at the first site of the second domain. For our analytical approach, we approximate $\beta_N$ as the bulk value of the second domain. A similar expression can be written for a $B$–$A$ boundary. To capture instabilities, we retain nonlinear terms of all orders in $\epsilon$ on the last site of the first domain. These terms dominate when a localized boundary mode at the kink is subject to inter-domain pressure. The effective model assumes the second domain remains unaffected by the first domain, i.e., that instabilities on opposite sides of the kink arise at different $\theta$ values in each domain.

The effective model~\eqref{eq:PreBogoliubovFluctuationHamiltonian} is written in the position basis, retaining nonlinear terms only at the last site of the domain. Transforming to the eigenmode basis that diagonalizes $H_{\rm fl}$ introduces 3- and 4-wave mixing among all eigenmodes due to these local nonlinearities, see Supporting Information. Similarly, the impact of inter-domain pressure on each mode scales with its localization on that site. As the pump progresses, the topological boundary mode localizes near the kink, therefore absorbing the most pressure (gain) from the other domain, and dominating the nonlinear dynamics. To capture this effectively, we approximate $\hat{a}_N \approx r \hat{d}_b$, where $\hat{d}_b$ annihilates a quasiparticle in the topological boundary mode of $H_{\rm fl}$. The coefficient $r$ is related to the boundary mode's amplitude on the last site and follows from the Bogoliubov transformation used to diagonalize $H_{\rm fl}$ [cf.~Eq.~\eqref{Eq:Rice-MelePart} and Supporting Information]. We thus identify the boundary-site dynamics with those of the boundary mode up to a proportionality factor.

Under this approximation, the bulk excitation spectrum remains unaffected. Only the boundary mode undergoes renormalization by inter-domain pressure and nonlinearities. The model $\tilde H_\mathrm{fl}$ then describes an effective energy for the boundary mode, whose renormalized eigenfrequency is given by the fluctuation frequency around the potential minimum. We describe this mode via a second mean-field ansatz $\hat{d}_b = \mathcal{D}_b + \tilde{\epsilon} \tilde{\hat{d}}_b$. Here, $\mathcal{D}_b$ describes the amplitude of the topological boundary mode minimizing the effective energy, $\tilde{\hat{d}}_b$ describes excitations around it, and $\tilde\epsilon$ is a small parameter. Expanding $\tilde{H}_{\rm fl}$ to second order in $\tilde{\epsilon}$ yields a new linearized Hamiltonian with the same bulk spectrum as $H_{\rm fl}$ but a renormalized boundary mode. A second Bogoliubov transformation then gives its renormalized energy
\begin{equation}
    \tilde{\omega}_b = \sqrt{\omega_b^2+\omega_b^{\phantom 2}\left(3r^3 \lambda\alpha_N^{\phantom 2}\mathcal{D}_b^{\phantom 2}+\frac{3}{2}r^4\lambda\mathcal{D}_b^2\right)}\,,
    \label{eq:ModifiedFrequency}
\end{equation}
where we set $\epsilon = \tilde\epsilon = 1$.

When $\tilde{\omega}_b$ becomes imaginary, the boundary mode and with it the many-body state becomes unstable, see Fig.~\ref{Fig:2}\textbf{c}. At this point, the system relaxes into a new configuration where the kink shifts by one site, marking a phase transition to another first-excited state, see Figs.~\ref{Fig:2}\textbf{a}, \textbf{b}. A similar analysis of the second domain under pressure from the first reveals a complementary instability region in $\theta$, see Fig.~\ref{Fig:2}\textbf{c}. The kink’s direction of motion is thus determined by which boundary mode destabilizes first during the pump. This mechanism is the origin of the kink displacement and represents the central result of this work. Notably, our simplified analytical prediction for the renormalized boundary mode energy agrees well with numerical simulations of the full system, see Supporting Information.

The displacement of the kink alters the termination between the two domains. As shown in Figs.~\ref{Fig:2}\textbf{d} and \textbf{e}, shifting the kink by one site is equivalent to removing a site from one domain and attaching it to the other. This exchange flips the sublattice configuration at the interface, which corresponds to shifting the pumping parameter $\theta$ by $\pi$ in both domains~\cite{kraus2012topological}.
Instead of recalculating the boundary state energies and instability conditions after each displacement, we incorporate this effect by applying this shift at the instability point, see Fig.~\ref{Fig:2}\textbf{c} (purple arrow). As a result, the system encounters the same instability region twice during a full pump cycle from $\theta = 0$ to $\theta = 2\pi$, causing the domain wall to shift by two sites (one full unit cell), see Fig.~\ref{Fig:2}\textbf{a}.
Reversing the direction of the pump reverses the sequence of instability crossings, and the kink moves in the opposite direction, see Fig.~\ref{Fig:2}\textbf{b}. 
\begin{figure}[!tb]
	\centering
	\includegraphics[width=1\columnwidth]{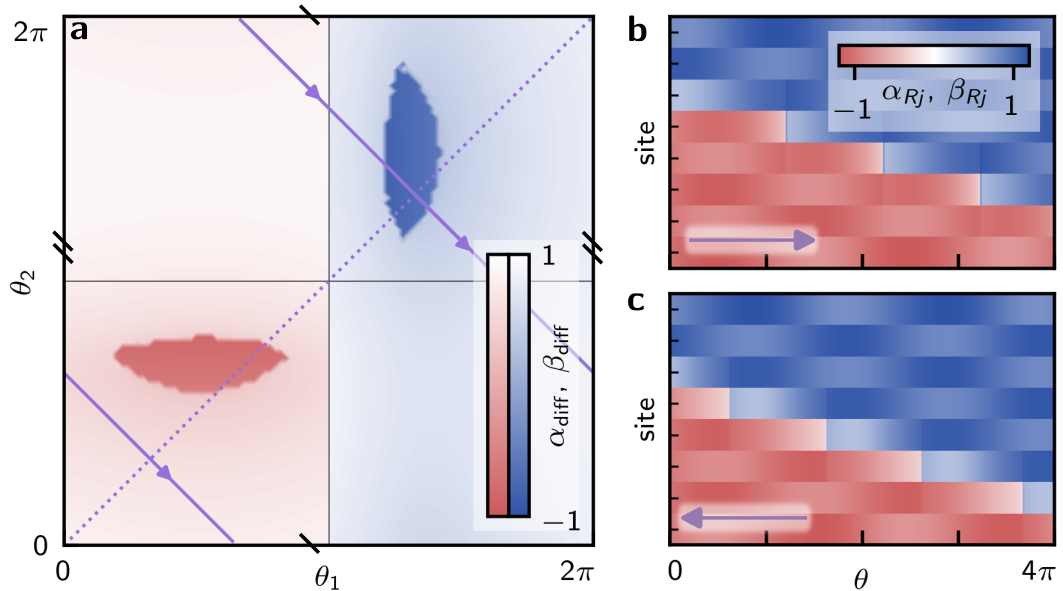}
	\caption{Generalized domain wall movement for two independent pumping parameters $\theta_1$ and $\theta_2$ for $\lambda=1$ and $\kappa=0.2$ in an 8-site chain.
    \textbf{a,} Instability phase diagram in the $\theta_1$-$\theta_2$ plane. Blue (red) regions mark parameters where the red (blue) domain is unstable as in Fig.~\ref{Fig:2}\textbf{f}. The purple, dotted line marks the pump cycle for $\theta_1=\theta_2$, the purple full line mark a pump cycle that only crosses the blue instability region.
    \textbf{b,} and \textbf{c,}, same as Fig.~\ref{Fig:1}\textbf{a} for a two pump parameter sweep along and against the direction of the arrows in \textbf{a}, respectively. $\theta_1$ ($\theta_2$) parametrizes the top (bottom) domain.}
	\label{Fig:3} 
\end{figure}

The quantized kink displacement depends on how often a boundary mode instability occurs during a pump cycle, which is determined by the trajectory in the $\Delta J$–$\Delta q$ space. In Fig.~\ref{Fig:2}\textbf{c}, we show the instability regions for the pump path in Fig.~\ref{Fig:1}\textbf{b}. To obtain a broader view, we examine stability across a wider parameter space; see Fig.~\ref{Fig:2}\textbf{f}. At each point, we initialize the domain wall configuration of Fig.~\ref{Fig:1}\textbf{a} and compute the corresponding two-domain state [cf.~Eq.~\eqref{eq:HamiltonianFullSystem}]. We then plot the ratio between the state's displacement at the site next to the kink ($\alpha_N$ or $\beta_N$) and in the bulk ($\alpha_0$ or $\beta_0$). The upper (lower) half shows $\beta_N/\beta_0$ ($\alpha_N/\alpha_0$) revealing an instability region for the blue (red) domain. These regions were also identified in the analytical model (Fig.~\ref{Fig:2}\textbf{c}). Crucially, unlike the Chern number dependent transport in the linear Rice–Mele model~\cite{Quantization_Thouless1983}, kink transport in the nonlinear case requires only entering an instability region. Nevertheless, the effect remains robust to disorder, see Supporting Information.

To illustrate the flexibility of our model, we generalize to a two-domain system with independent pumping parameters. We consider a configuration similar to Fig.~\ref{Fig:1}\textbf{a}, where the red (blue) domain is controlled by $\theta_1$ ($\theta_2$), and the inter-domain coupling is fixed at $\kappa$. Each domain is parameterized by $\Delta q_n = \sin(\theta_n)$ and $\Delta J_n = \cos(\theta_n)$ with $n = 1, 2$. As in the single-parameter case, we expect kink displacement triggered by boundary mode instabilities.
We compute the instability diagram of this extended system in Fig.~\ref{Fig:3}\textbf{a}, following the procedure in Fig.~\ref{Fig:2}\textbf{f}. Two distinct instability regions appear, corresponding to opposite directions of kink motion. When $\theta_1 = \theta_2$, the system reduces to the single-parameter pump, differing only by fixed inter-domain coupling. Hence, both types of instabilities may occur, depending on pump direction. 
However, tuning $\theta_1$ and $\theta_2$ independently allows biasing of the system to encounter only one instability region, regardless of the pumping direction. The kink then moves unidirectionally, see Figs.~\ref{Fig:3}\textbf{b} and \textbf{c}, computed similarly to Figs.~\ref{Fig:2}\textbf{a},\textbf{b}. This directional locking demonstrates how the nonlinear mechanism differs fundamentally from linear topological pumps, which are governed by Chern numbers. Instead, kink displacement in our model arises from localized instabilities shaped by nonlinear interactions.
\begin{figure}[!tb]
	\centering
	\includegraphics[width=1\columnwidth]{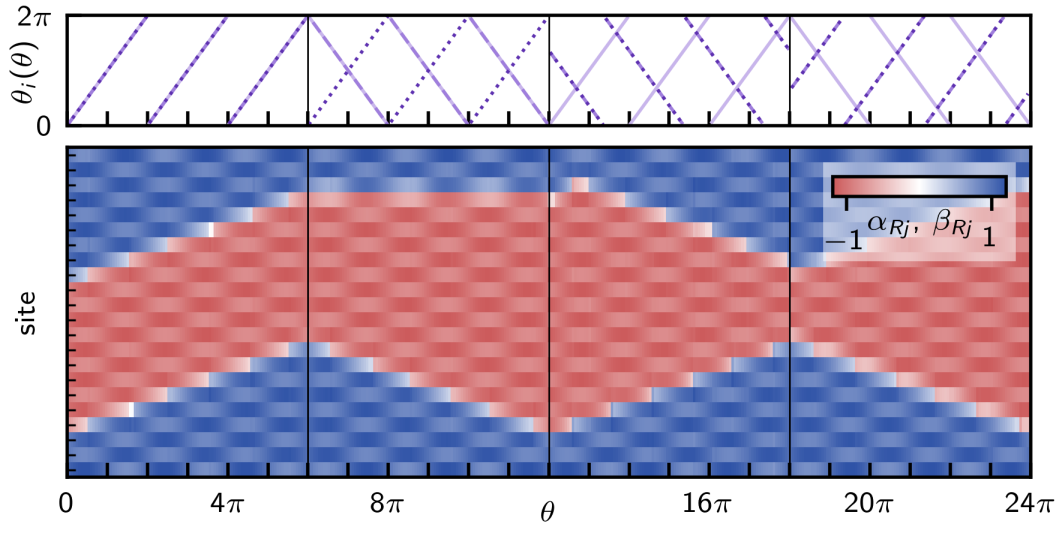}
	\caption{Soliton dynamics for a three-domain scenario using three independent pumping parameters $\theta_1$, $\theta_2$ and $\theta_3$, each parameterizing one domain in a 22-site chain (bottom to top in the bottom panel) with $\lambda=1$ and $\kappa=0.2$. Top: Different variations of pumping parameters, where $\theta_1$, $\theta_2$ and $\theta_3$ are represented by full, dashed and dotted purple lines, respectively. Bottom: Movement of domains for 12 pump cycles while varying the pumping parameters according to the top panel for the same parameters as in Fig.~\ref{Fig:2}.}
	\label{Fig:4} 
\end{figure}
%

We now extend the setup to a three-domain configuration, containing two kinks and a central domain that can be understood to act as a soliton. Each domain is controlled by an independent pumping parameter, $\theta_n$, with $\Delta q_n = \sin(\theta_n)$ and $\Delta J_n =\cos(\theta_n)$ for $n = 1,2,3$. The couplings between the domains are kept fixed at $\kappa$. As long as the two kinks remain well separated, the instability diagrams at the domain boundaries are equivalent to those of the two-domain case shown in Fig.~\ref{Fig:3}\textbf{a}.
The extended control offered by three pumping parameters allows us to freely manipulate the shape and motion of the central domain, as illustrated in Fig.~\ref{Fig:4}. Depending on the chosen pumping protocol, the soliton-like domain can be moved, expanded, compressed, or deformed asymmetrically. Remarkably, even a uniform protocol with $\theta_1 = \theta_2 = \theta_3$ is sufficient to drive the central domain across the system.
The motion of the soliton results from independent boundary instabilities at each of its kinks. As before, this mechanism cannot be described by the Chern number of any single domain. It highlights the fundamentally nonlinear and non-adiabatic nature of kink transport in our system.

In conclusion, our work introduces a quantized pumping phenomenon for propagating a nonlinear kink and interprets nonlinear topological pumping from the perspective of the system's boundary. The quantized kink motion arises from instabilities of topologically protected boundary modes of the underlying linear model. These instabilities result from pressure exerted by the nonlinear kink, modifying the boundary mode's energy. Evidently, the topological invariant of the linear model, the Chern number, cannot describe the kink's movement, as the latter depends solely on the number of instabilities reached throughout a pump cycle. Our model applies broadly to nonlinear bosonic platforms, including photonics, ultracold atoms, and mechanical metamaterials. Future work will explore how this framework explains recent findings, such as fractional nonlinear pumping \cite{jurgensen2023quantized,jürgensen2025multibandfractionalthoulesspumps,Tao2025NonlinearityInducedFractionalTopoPump,Fleischhauer2025FractionalTopoPump}, nonlinearity-induced pumping~\cite{Tao2024nonlinearityinducedthoulesspumpingsolitons,Tao2025NonlinearityInducedFractionalTopoPump} and the non-adiabatic breakdown of pumping~\cite{Tuloup2023InstabilitiesInPumpCycle} from a microscopic perspective.

\section*{Acknowledgements}
We thank fruitful discussions with N.~Pernet, S.~Ravets, J.~Bloch, M.~S. Garcia, and J.~del Pino. We acknowledge funding from the Deutsche Forschungsgemeinschaft (DFG) via project numbers 449653034, 521530974, and 545605411, as well as through SFB1432 on project number 425217212. We furthermore acknowledge funding from the Swiss National Science Foundation (SNSF) through the Sinergia Grant No.~CRSII5\_206008/1 and NCCR SPIN.

\bibliography{bibliography.bib}


\clearpage
\onecolumngrid

\setcounter{equation}{0}
\setcounter{figure}{0}
\setcounter{table}{0}
\setcounter{page}{1}
\setcounter{section}{0}
\makeatletter
\renewcommand{\theequation}{S\arabic{equation}}
\renewcommand{\thesection}{S\arabic{section}}
\renewcommand{\thefigure}{S\arabic{figure}}
\renewcommand{\bibnumfmt}[1]{[S#1]}

\begin{center}
\textbf{\large Supporting Information: Quantized nonlinear kink movement through topological boundary state instabilities}
\end{center}

\section{Chern number of the linarized model}

\begin{figure}[!htb]
	\centering
	\includegraphics[width=0.5\columnwidth]{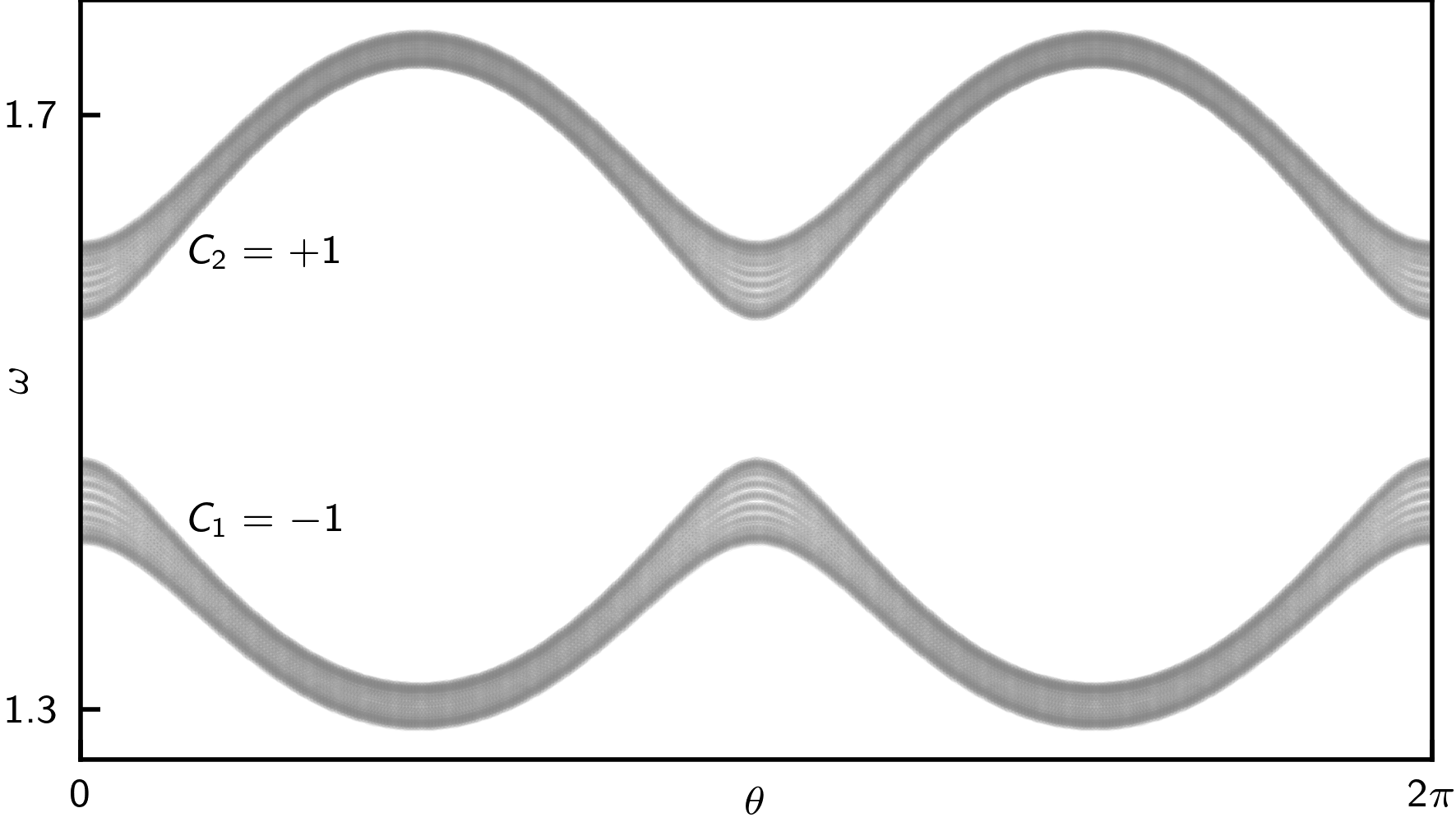}
	\caption{Bandstructure calculated via Eq.~\eqref{SIeq:MomentumSpaceHamiltonian} for same parameters as in Fig.~\ref{Fig:2}. The Chern number $C_n$ of the upper (lower) band is $+1$ ($-1$).}
	\label{Fig:ChernNumber} 
\end{figure}

The fluctuation Hamiltonian of the linearized system is $H_\mathrm{fl}=H_\mathrm{RM}+H_\mathrm{sq}$. $H_\mathrm{RM}$ reads
\begin{equation}
    H_\mathrm{RM} = \sum_j \bigg[\delta_A(\theta) \ah_j^\dagger \ah_j^{\phantom \dagger} + \delta_B(\theta) \bh_j^\dagger \bh_j^{\phantom \dagger}
    -\frac{J_1(\theta)}{4}\ah_j^\dagger \bh_j^{\phantom \dagger} - \frac{J_2(\theta)}{4}\bh_j^\dagger \ah_{j+1}^{\phantom \dagger} +h.c.\bigg]\, ,
\end{equation}
cf. Eq.~(2) in the main text. The squeezing part of the Hamiltonian takes a similar form with
\begin{equation}
    H_\mathrm{sq}=\sum_j (\delta_A(\theta)-\frac{1}{2})(a_j^\dagger a_j^\dagger + a_j a_j)+(\delta_B(\theta)-\frac{1}{2})(b_j^\dagger b_j^\dagger + b_j b_j)-\frac{J_1(\theta)}{4}(a_j b_j + a_j^\dagger b_j^\dagger)-\frac{J_2(\theta)}{4}(a_{j+1} b_j+a_{j+1}^\dagger b_j^\dagger)
\end{equation}
where $\delta_A(\theta)$ and $\delta_B(\theta)$ are defined as in the main text. For the calculation of the Chern number of this linear system, we move to reciprocal space. We define $a_j=\sum_k e^{ijk} a_k$ and $b_j=\sum_k e^{ijk} b_k$ for the quasi momentum $k$ and arrive at
\begin{align}
H_\mathrm{fl}(\theta)=\sum_k \mathbf{a}_k^\dagger \begin{pmatrix}
    \delta_A & -\frac{1}{4}(J_1+J_2 e^{-ik}) & \delta_A-\frac{1}{2} & -\frac{1}{4}(J_1+J_2 e^{-ik})\\
     -\frac{1}{4}(J_1+J_2 e^{ik})& \delta _B & -\frac{1}{4}(J_1+J_2 e^{ik}) & \delta_B-\frac{1}{2} \\
     \delta_A-\frac{1}{2}&-\frac{1}{4}(J_1+J_2 e^{-ik}) &\delta_A & -\frac{1}{4}(J_1+J_2 e^{ik})\\
     -\frac{1}{4}(J_1+J_2 e^{ik})& \delta_B-\frac{1}{2} & -\frac{1}{4}(J_1+J_2 e^{-ik}) & \delta_B
\end{pmatrix}    
    \mathbf{a}_k
    \label{SIeq:MomentumSpaceHamiltonian}
\end{align}
with $\mathbf{a}_k=(a_k, b_k, a_{-k}^\dagger,b_{-k}^\dagger)^T$.
The Chern number $C_n$ of the $n$-th band then reads
\begin{align}
    C_n = \frac{1}{2\pi i}\int_0^{2\pi}d\theta\int_0^{2\pi}dk\,(\partial_\theta A_k-\partial_k A_\theta)\, ,
\end{align}
where $A_\mu = \langle n(k, \theta)|\partial_\mu|n(k,\theta)\rangle$ is the Berry connection with $\mu\in\{\theta,k\}$. Here, $|n(k,\theta)\rangle$ denotes the normalized eigenmodes of $H_\mathrm{fl}$.
Here, we interpret $\theta$ as an artificial quasi momentum and the chain therefore as a $(1+1)$ dimensional system. We calculate the Chern number numerically \cite{Fukui2005ChernNumberNumerical} for the same parameters and pumping trajectory as in Fig.~2\textbf{c} in the main text, see Fig.~\ref{Fig:ChernNumber}. 
This yields a Chern number of $+1$ ($-1$) for the upper (lower) band, see Fig.~\ref{Fig:ChernNumber}, confirming the nontrivial topology of the linearized model of one domain. 

\section{Details on the modified boundary mode energy calculation}
In this section, we show the details of the derivation of Eq.~(4) in the main text. For this, we follow the steps described there.
We start with the Hamiltonian in Eq.~(1) in the main text
\begin{align}
    H(\theta)=\sum_j\bigg[\sum_{l=A,B} \Big(\frac{1}{2}\hat p_{j,l}^2-\frac{q_l(\theta)}{2}\hat x_{j,l}^2+\frac{\lambda}{4} \hat x_{j,l}^4\Big)
        +\frac{J_1(\theta)}{2}\left(\hat x_{j,A}-\hat x_{j,B}\right)^2
        +\frac{J_2(\theta)}{2}\left(\hat x_{j,B}-\hat x_{j+1,A}\right)^2\bigg]\, .
\end{align}
We first insert the mean field ansatz $\hat x_{j,A} = 1/\sqrt{2}\left((\alpha_j+\alpha_j^*)+\epsilon(\hat a_j +\hat a _j^\dagger)\right)$ and $\hat p_{j,A} = i/\sqrt{2}\left((\alpha_j^*-\alpha_j)+\epsilon(\hat a _j^\dagger-\hat a_j)\right)$ (analogous for $\hat{x}_{j,B}$ and $\hat{p}_{j,B}$ with $\alpha_j,\ \hat a_j\rightarrow \beta_j, \, \hat b_j$) into $H(\theta)$. Then, we split the Hamiltonian $H=H_\mathrm{MF}+H_\mathrm{fl}$ into mean-field part $H_\mathrm{MF}$ and fluctuation part $H_\mathrm{fl}$. The mean field part reads
\begin{align}
    H_{MF} =& \sum_j -\frac{1}{4}(\alpha_{j}^*-\alpha_{j})^2-\frac{q_A(\theta)}{4}(\alpha_{j}^*+\alpha_{j})^2+\frac{\lambda}{16} (\alpha_{j}^*+\alpha_{j})^4-\frac{1}{4}(\beta_{j}^*-\beta_{j})^2-\frac{q_B(\theta)}{4}(\beta_{j}^*+\beta_{j})^2+\frac{\lambda}{16} (\beta_{j}^*+\beta_{j})^4
    \\
    &+\frac{J_1(\theta)}{4}\left((\alpha_{j}^*-\alpha_{j})-(\beta_{j}^*+\beta_{j})\right)^2
    +\frac{J_2(\theta)}{4}\left((\beta_{j}^*+\beta_{j})-(\alpha_{j+1}^*+\alpha_{j+1})\right)^2\, .\nonumber
\end{align}
We first consider the situation with only one domain.
When choosing $\alpha_j$ and $\beta_j$ as stable states of $H_\mathrm{MF}$ (we choose the one-domain state), the terms of order $\epsilon$ drop out in $H_\mathrm{FL}$. We then get
\begin{align}
    H_\mathrm{FL} =& \sum_j -\frac{\epsilon^2}{4}(\ah_{j}^\dagger-\ah_{j})^2-\frac{\epsilon^2 q_A(\theta)}{4}(\ah_{j}^\dagger+\ah_{j})^2+\frac{\lambda}{16}(6\epsilon^2(\alpha_j^*+\alpha_j)^2(\ah_{j}^\dagger+\ah_{j})^2 + 4\epsilon^3(\alpha_j^*+\alpha_j)(\ah_{j}^\dagger+\ah_{j})^3 + \epsilon^4(\ah_{j}^\dagger+\ah_{j})^4) \nonumber
    \\
    &-\frac{\epsilon^2}{4}(\bh_{j}^\dagger-\bh_{j})^2-\frac{q_B(\theta)}{4}(\bh_{j}^\dagger+\bh_{j})^2+\frac{\lambda}{16}(6\epsilon^2(\beta_j^*+\beta_j)^2(\bh_{j}^\dagger+\bh_{j})^2 + 4\epsilon^3(\beta_j^*+\beta_j)(\bh_{j}^\dagger+\bh_{j})^3 + \epsilon^4(\bh_{j}^\dagger+\bh_{j})^4)\nonumber
    \\
    &+\frac{\epsilon^2 J_1(\theta)}{4}\left((\ah_{j}^\dagger-\ah_{j})-(\bh_{j}^\dagger+\bh_{j})\right)^2
    +\frac{\epsilon^2 J_2(\theta)}{4}\left((\bh_{j}^\dagger+\bh_{j})-(\ah_{j+1}^\dagger+\ah_{j+1})\right)^2\, .
    \label{SIeq:AllFluctuationsHamiltonian}
\end{align}
Retaining only terms to second order in $\epsilon$ in
Eq.~\eqref{SIeq:AllFluctuationsHamiltonian}, leads to $H_\mathrm{fl}$ in the main text. We can therefore rewrite $H_\mathrm{FL}$ as
\begin{align}
    H_{\mathrm{FL}} = \epsilon^2 H_{\mathrm{fl}} +\sum_j \frac{\lambda}{16}\Big( 4\epsilon^3(\alpha_j^*+\alpha_j)(\ah_{j}^\dagger+\ah_{j})^3 + \epsilon^4(\ah_{j}^\dagger+\ah_{j})^4
    +4\epsilon^3(\beta_j^*+\beta_j)(\bh_{j}^\dagger+\bh_{j})^3 + \epsilon^4(\bh_{j}^\dagger+\bh_{j})^4\Big)\,.
\end{align}
We now consider a finite chain and assume that the mean-field solutions do not change at the end of the chain. Then, we diagonalize $H_\mathrm{fl}$ via a Bogoliubov transformation \cite{xiao2009theorytransformationdiagonalizationquadratic,Colpa1978BogoliubovTrafo}. We therefore rewrite $H_\mathrm{fl}=\psi^\dagger \mathbf{H}_\mathrm{fl}\psi$ with $\psi^T=((\vec{a})^T, \vec{a}^\dagger)$, $\vec{a}=(...,\ah_j, \bh_j, \ah_{j+1}...)^T$ and the matrix $\mathbf{H}_\mathrm{fl}$. We then use the Bogoliubov transformation matrix $\mathbf{T}$ to find the eigenmodes of the Hamiltonian
\begin{align}
    H_\mathrm{fl}=\psi^\dagger(\mathbf{T^\dagger})^{-1}\mathbf{T}^\dagger \mathbf{H}_{\mathrm{fl}} \mathbf{T}\mathbf{T}^{-1}\psi=\sum_{j}\omega_j\hat d_j^\dagger \hat d_j
\end{align}
with
\begin{align}
    \hat d_j=(\mathbf{T}^{-1}\psi)_j=(\phi)_j=((\vec{d})^T, \vec{d}^\dagger)_j\, ,
    \label{eq:BogoliubovTransformation}
\end{align}
where $\vec{d}=(...,\hat d_j, \hat{d}_{j+1},...)^T$. Here, $\hat d_j$ is the annihilation operator of the $j$-th eigenmode of $H_\mathrm{fl}$ with eigenenergy $\omega_j$. Note that, here, the Bogoliubov transformation matrix $\mathbf{T}$ is chosen such that $\hat d_j$ and $\hat d_j^\dagger$ obey bosonic commutation relations. This leads to the band structure in Fig.~2\textbf{c} in the main text with two boundary modes crossing the gap.

We now consider the influence of a second domain. For this, we only take into account one site of the second domain that only couples via its mean field solution to the first domain. The last site of the first domain is the $A$ site in the $N$-th unit cell. We include this coupling into the fluctuation Hamiltonian via
\begin{align}
    H_{FL} = &\epsilon^2 H_{fl} +\sum_j \frac{\lambda}{16}\Big( 4\epsilon^3(\alpha_j^*+\alpha_j)(\ah_{j}^\dagger+\ah_{j})^3 + \epsilon^4(\ah_{j}^\dagger+\ah_{j})^4
    +4\epsilon^3(\beta_j^*+\beta_j)(\bh_{j}^\dagger+\bh_{j})^3 + \epsilon^4(\bh_{j}^\dagger+\bh_{j})^4\Big) \\
    &+ \frac{\epsilon J_1(\theta)}{2}((\beta_{N-1}^\dagger+\beta_{N-1})-(\beta_{N}^\dagger+\beta_{N}))(\ah_N+\ah_N^\dagger)\,.\nonumber
\end{align}
The $\beta_{N-1}^\dagger+\beta_{N-1}$ compensates for the mean-field solutions assuming that site $N, \, B$ is still part of the first domain, while the $\beta_{N}^\dagger+\beta_{N}$ term is the additional strain by the boundary.
Because of the symmetry of the model, the relation $(\beta_{N-1}^\dagger+\beta_{N-1})=-(\beta_{N}^\dagger+\beta_{N})$ holds. We now furthermore assume, that the influence of the second domain is highly localized on the last site of the first domain. We therefore neglect all terms with higher order than $\epsilon^2$ from other sites and arrive at Eq.~(3) in the main text
\begin{align}
\tilde{H}_\mathrm{fl}=&
    \epsilon^2 H_\mathrm{fl}
    +\frac{\epsilon^3\lambda}{4}(\alpha_N+\alpha_N^*) (\ah_N^{\phantom \dagger}+\ah_N^\dagger)^3 + \frac{\epsilon^4\lambda}{16}(\ah_N^{\phantom \dagger}+\ah_N^\dagger)^4 -\epsilon J_1(\theta)(\beta_{N}^{\phantom *}+\beta_{N}^*)(\ah_N^{\phantom \dagger}+\ah_N^\dagger)\,.
    \label{eq}
\end{align}
As $\epsilon$ fullfiled its role as expansion parameter, we set $\epsilon=1$.
The Bogoliubov transformation [Eq.~\eqref{eq:BogoliubovTransformation}] for $\ah_N +\ah_N^\dagger$ is 
\begin{align}
    (\ah_N+\ah_N^\dagger)=(\vec{e}_N+\vec{e}_{2N})^T\mathbf{T}\begin{pmatrix}
        \vec{d}\\
        (\vec{d}^\dagger)^T
    \end{pmatrix}\, ,
    \label{eq:BogoliubovTransformSum}
\end{align}
where $\vec{e}_j$ is the vector of length $2N$ with $0$ everywhere, but a $1$ on the $j$-th entry. However, as explained in the main text, insering this Bogoliubov transformation into Eq.~\eqref{eq:PreBogoliubovFluctuationHamiltonian} leads to 3- and 4-wave mixing terms in the Hamiltonian. To continue with a simple analytical expression for the eigenenergy of the boundary mode, we approximate the Bogoliubov transform in Eq.~\eqref{eq:BogoliubovTransformSum}: since the boundary mode is highly localized, we approximate its corresponding annihilation and creation operators $\ddh_b$ and $\ddh_b^\dagger$ as proportional to the operators of the last site $\ah_N$ and $\ah_N^\dagger$:
\begin{equation}
(\ah_N + \ah_N^\dagger)
\approx
\frac{\mathbf{T}_{N,b}+\mathbf{T}_{2N,b}}{\sqrt{c}}
(\ddh_b + \ddh_b^\dagger)
= r (\ddh_b + \ddh_b^\dagger)\, ,
\label{eq:ApproximatedBogoliubov}
\end{equation}
with
$c=(\mathbf{T}_{N,b}+\mathbf{T}_{2N,b})(\mathbf{T}_{b,N}^{-1}+\mathbf{T}_{b+N,N}^{-1})$.
We introduced the scaling factor $c$, which ensures that the approximated Bogoliubov transform in Eq.~\eqref{eq:ApproximatedBogoliubov} is still acting as a standard Bogoliubov transform. Inserting Eq.~\eqref{eq:ApproximatedBogoliubov} into Eq.~(3) in the main text, we arrive at
\begin{equation}
    \begin{aligned}
        \tilde{H}_{\mathrm{fl}}=& H_{\mathrm{fl}}
        +\frac{\epsilon^3\lambda}{4}(\alpha_N+\alpha_N^*) r^3 (\ddh_b+\ddh_b^\dagger)^3 
        + \frac{\epsilon^4\lambda}{16} r^4 (\ddh_b+\ddh_b^\dagger)^4-\epsilon J_1(\theta)r(\beta_{N}+\beta_{N}^*)  (\ddh_b+\ddh_b^\dagger)\,.
    \end{aligned}
\end{equation}
From here, we can now employ a second mean field ansatz as shown in the main text to arrive at the modified fluctuation energy as in Eq.~(4) in the main text.

\section{Linearized fluctuation spectrum of the system obtained through exact diagonalization}
\begin{figure}[!htb]
	\centering
	\includegraphics[width=1\columnwidth]{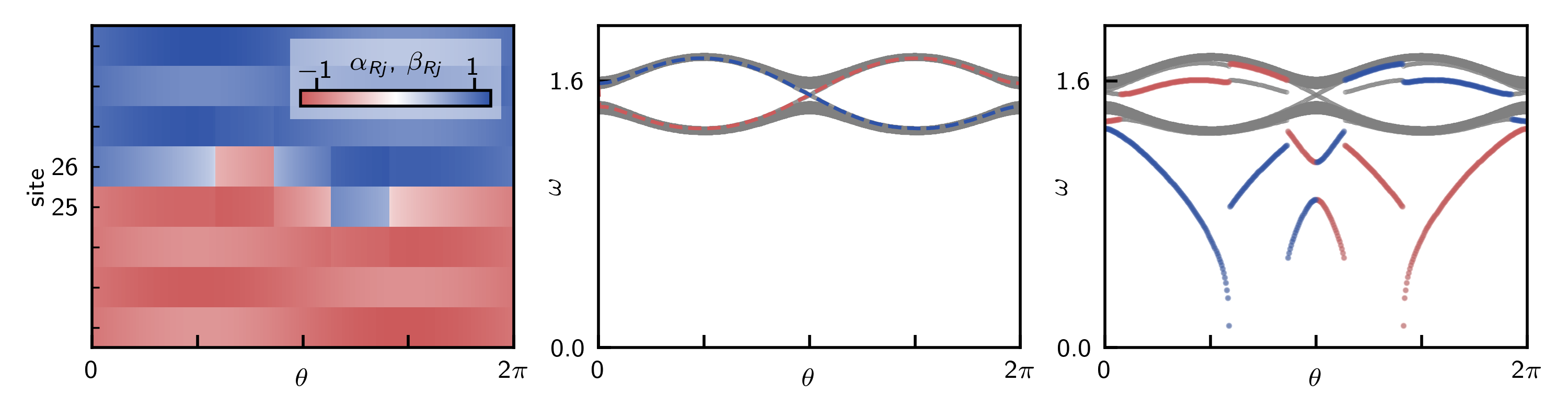}
	\caption{Exact diagonalization for the eigenmodes of the chain for the same parameters as in Fig.~2 in the main text. 
    \textbf{a,} Zoom in on the nonlinear kink throughout one pump cycle. For each step in the pump cycle, the system of 50 sites is initialized in the configuration depicted in Fig.~1\textbf{a} in the main text with a nonlinear kink between sites $25$ and $26$ of the chain and then is let to evolve to a stable state. This is different to Fig.~2 in the main text, as we do not take the stable state for the previous $\theta$ as a seed. The kink therefore does not move through the chain as in Fig.~2\textbf{a} and \textbf{b} in the main text.
    \textbf{b,} Numerical band structure of an open chain without a kink. boundary modes localized at site $0$ and $50$ cross the band gap at $\theta=\pi$.
    \textbf{c,} Band structure of the chain with the nonlinear kink depicted in \textbf{a} calculated through exact diagonalization. The modes most localized on site $25$ ($26$) are marked in red (blue).}
	\label{Fig:ED} 
\end{figure}

To verify the validity of the approach used for calculating the instabilities of the boundary modes in the main text, we can compare the analytical calculation of the main text to a full diagonalization of the system. We first discretize $\theta\in[0,2\pi)$. For each discretized value of $\theta$, we initialize a chain in the initial condition depicted in Fig.~1\textbf{a} in the main text with the nonlinear kink in the middle of the chain. We then evolve the system to a stable mean field steady state (see Fig.~\ref{Fig:ED}\textbf{a}), similar as for Fig.~2\textbf{a} and \textbf{b} in the main text. Note, that here we initialize the system for each $\theta$ in the state which has a kink between site $25$ and $26$, regardless of the state the system ended up for the previous value of $\theta$. This procedure therefore differs from the one used in Fig.~2\textbf{a} and \textbf{b}, as the kink does not move through the system when pumping. We then use the resulting steady state to calculate the Bogoliubov fluctuation spectrum via Eq.~(2) in the main text. The main difference to the procedure performed in the main text is that the fluctuation Hamiltonian in the main text did not include the influence of the kink on the mean field solution of the domains.
The bandstructure calculated using exact diagonalization (Fig~\ref{Fig:ED}\textbf{c}) shows, the same behavior of boundary modes as predicted analytically in the main text. The movement of the kink coincides with boundary modes approaching $\omega=0$. The main difference to Fig.~\ref{Fig:2}\textbf{a} is that the exact diagonalization also shows hybridization between the detuned boundary modes for $\theta\approx\pi$. This effect was ignored in the effective energy calculation in the main text, as we treated the two boundary modes independently of one another. Since the hybridization only occurs in a relatively small section throughout the pump, it does not impact the occurence of instabilities in this case.

\section{Robustness against disorder}
\begin{figure}[!htb]
	\centering
	\includegraphics[width=0.5\columnwidth]{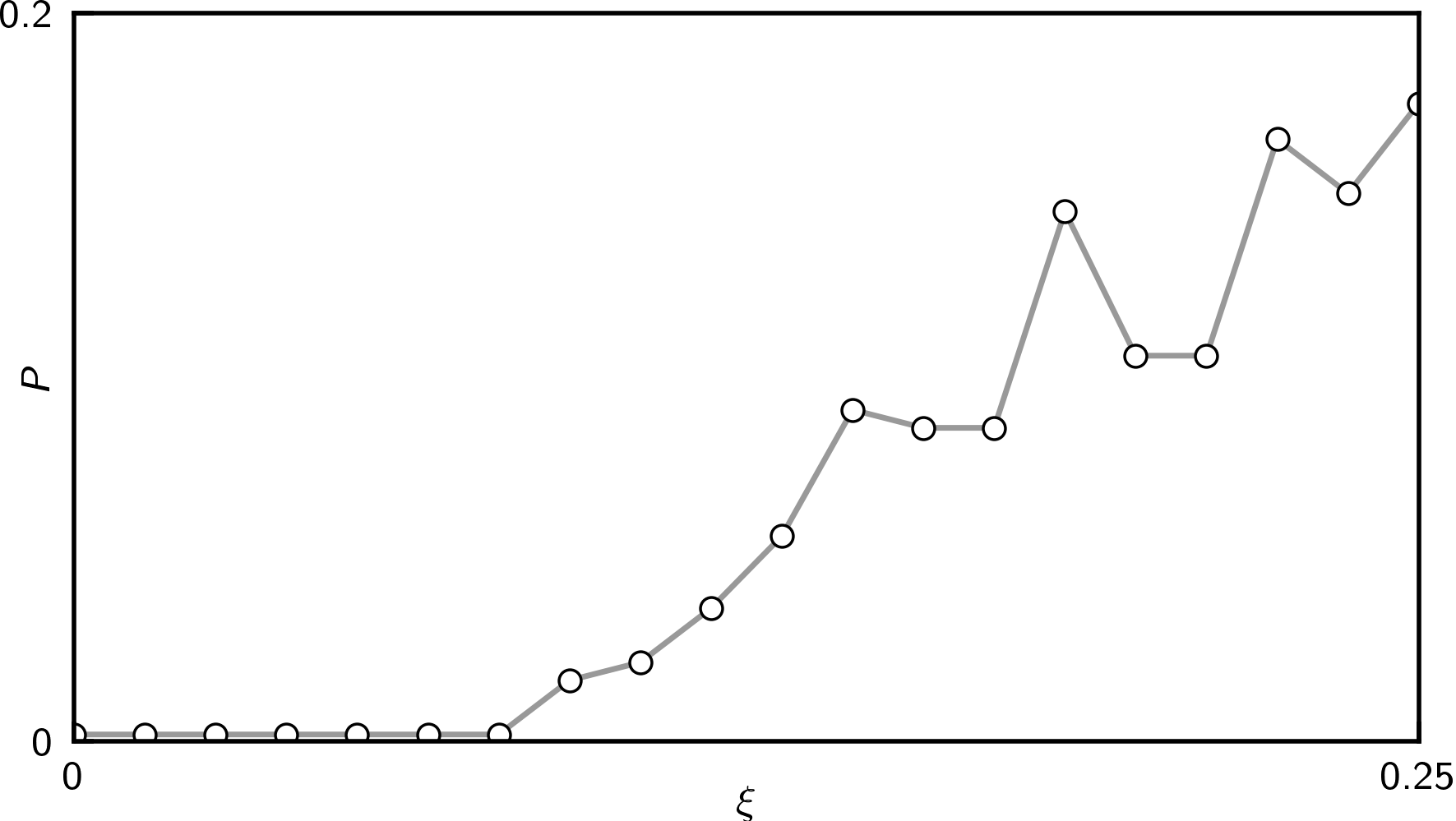}
	\caption{Robustness of the kink movement to random disorder. Parameters are the same as in Fig.~2 in the main text with $\tilde{q}_l(\theta)$ instead of $q_l(\theta)$. We sample $200$ random initializations for each value of $\xi$.}
	\label{Fig:Disorder} 
\end{figure}

A remarkable property of topological effects is their robustness against disorder. To probe the robustness of kink movement through the system, we initialize the system with one domain wall (see Fig.~1\textbf{a} in the main text) and additional disorder of the quadratic terms of the double-well potentials $\tilde{q}_l(\theta) = q_l(\theta)(1+\xi \mathcal{N}_l)$. Here, $\xi$ is the disorder strength and $\mathcal{N}_l$ is a random variable sampled from a standard normal distribution. We then numerically evolve $200$ initializations for different values of $\xi$ of the system from $\theta=0$ to $\theta=\pi$ as in Fig.~2\textbf{a} and \textbf{b} in the main text and calculate the percentage $P$ of initializations where the kink did not shift its position form the initial configuration, see Fig.~\ref{Fig:Disorder}.
The nonlinear kink movement through instabilities shows strong stability against disorder. For small disorders, the nonlinear kink movement is fully unaffected by the presence of disorder. Then, for higher disorder strengths ($\xi\approx 0.08$) a finite probability sets in and the disorder can block the kink from moving.

\end{document}